\documentclass[iop,apjl]{emulateapj}

\usepackage{amsmath}
\usepackage{color}
\newcommand{\gps}{\ensuremath{g_{\rm P1}}}
\newcommand{\rps}{\ensuremath{r_{\rm P1}}}
\newcommand{\ips}{\ensuremath{i_{\rm P1}}}

\shorttitle{Sagittarius Stream in PS1}
\shortauthors{Slater et al.}

\begin{document}

\title{A Pan-STARRS1 View of the Bifurcated Sagittarius Stream}

\author{C. T. Slater\altaffilmark{1},
E. F. Bell\altaffilmark{1},
E. F. Schlafly\altaffilmark{2},
M. Juri\'{c}\altaffilmark{3,4,5,6},
N.F. Martin\altaffilmark{7,8},
H.-W. Rix\altaffilmark{8},
E. J. Bernard\altaffilmark{9},
W. S. Burgett\altaffilmark{10},
K. C. Chambers\altaffilmark{10},
D. P. Finkbeiner\altaffilmark{3},
B. Goldman\altaffilmark{8},
N. Kaiser\altaffilmark{10},
E. A. Magnier\altaffilmark{10},
E. P. Morganson\altaffilmark{8},
P. A. Price\altaffilmark{11}, and
J. L. Tonry\altaffilmark{10} 
}
\affil{}

\altaffiltext{1}{Department of Astronomy, University of Michigan,
    500 Church St., Ann Arbor, MI 48109; ctslater@umich.edu}
\altaffiltext{2}{Department of Physics, Harvard University, 17 Oxford Street,
Cambridge MA 02138}
\altaffiltext{3}{Harvard-Smithsonian Center for Astrophysics, 60 Garden Street,
Cambridge, MA 02138}
\altaffiltext{4}{Hubble Fellow}
\altaffiltext{5}{Steward Observatory, University of Arizona, Tucson, AZ 85121}
\altaffiltext{6}{LSST Corporation, 933 N. Cherry Ave., Tucson, AZ 85721}
\altaffiltext{7}{Observatoire astronomique de Strasbourg, Universit\'e de
Strasbourg, CNRS, UMR 7550, 11 rue de l’Universit\'e, F-67000 Strasbourg,
France}
\altaffiltext{8}{Max-Planck-Institut f\"{u}r Astronomie, K\"{o}nigstuhl 17,
D-69117 Heidelberg, Germany}
\altaffiltext{9}{SUPA, Institute for Astronomy, University of Edinburgh, Royal
Observatory, Blackford Hill, Edinburgh EH9 3HJ}
\altaffiltext{10}{Institute for Astronomy, University of Hawaii at Manoa,
Honolulu, HI 96822, USA}
\altaffiltext{11}{Princeton University Observatory, 4 Ivy Lane, Peyton Hall,
Princeton University, Princeton, NJ 08544}

\begin{abstract}
We use data from the Pan-STARRS1 survey to present a panoramic view of the
Sagittarius tidal stream in the southern Galactic hemisphere. As a result of the
extensive sky coverage of Pan-STARRS1, the southern stream is visible along more
than $60^\circ$ of its orbit, nearly double the length seen by the SDSS. The
recently discovered southern bifurcation of the stream is also apparent,
with the fainter branch of the stream visible over at least $30^\circ$. Using a
combination of fitting both the main sequence turn-off and the red clump, we
measure the distance to both arms of the stream in the south. We find that the
distances to the bright arm of the stream agree very well with the N-body models
of \citet{law10}. We also find that the faint arm lies ${\sim} 5$ kpc closer to
the Sun than the bright arm, similar to the behavior seen in the northern
hemisphere. 
\end{abstract}


\section{Introduction}

The past decade has seen a tremendous growth in the amount of known tidal
substructure in the Galactic halo \citep{ibata01,newberg02,
yanny03,martin04,grillmair06a,grillmair06b,belokurov07a,belokurov07b,juric08}.
One of the most prominent and well-studied tidal streams in the halo is the
Sagittarius stream. A wide variety of tracers have been used to study the
stream, including main sequence turn-off (MSTO) stars \citep{belokurov07b}, blue
horizontal branch (BHB) stars \citep{niederste-ostholt10,ruhland11}, A stars
\citep{yanny00,martinez-delgado01,martinez-delgado04}, M giants
\citep{majewski03}, RR Lyrae \citep{vivas01,vivas05,watkins09} and red clump
(RC) stars \citep{bellazzini06,correnti10}. These studies have all revealed that
the Sgr stream is exceedingly complex and highly structured. The northern
``bifurcation'', where the stream appears to split into two parallel streams,
remains unexplained. Multiple possible causes of the bifurcation have been
proposed \citep[e.g.,][]{fellhauer06,penarrubia10} but no explanation so far has
proved satisfactory \citep[e.g.,][]{penarrubia11}. 

Recent work by \citet{koposov12} using the SDSS has shown that the Sgr stream is
more complex than previously believed, with a bifurcation in the stream in the
southern Galactic hemisphere as well as in the north. This has only added to the
challenge of reconciling the diverse properties of the stream. It also changes
the range of possible explanations, e.g., it is possible that each of the two
northern streams comes from a different progenitor.  As a result of all these
features of the stream itself, along with uncertainties about the gravitational
potential of the Galaxy \citep{helmi04,law05,johnston05,fellhauer06}, the Sgr
stream has proven to be a serious challenge to any model that attempts to
describe it. The model of \citet{law10} has been able to generally reproduce
many of the features of the stream (assuming a triaxial Galactic potential), but
the bifurcations are not yet accounted for.

These modeling challenges underscore the importance of obtaining a complete set
of observations of the stream, including data from both the northern and southern
Galactic hemispheres. In this work we focus on the southern component of the
stream, using data from the Pan-STARRS1 project. Pan-STARRS's extensive sky
coverage allows for a comprehensive view over approximately $60^\circ$ along the
southern stream, yielding a picture in the south comparable to that available in
the north.

In this work we extend the coverage of the southern Sgr stream seen by
\citet{koposov12} in SDSS, nearly doubling the length of the stream observed. We
measure distances along the entire observed southern stream, using red clump
(RC) stars and a direct calibration of the RC absolute magnitude from the Sgr
dwarf itself.  The red clump is an excellent distance indicator, since it
occupies a narrow range of absolute magnitudes and is only weakly dependent on
the parameters of the stellar population. However, the red clump is a relatively
weak feature, and in faint systems it can be difficult to unambiguously identify
without additional information. This is particularly true at low galactic
latitudes where the large number of disk stars increases the difficulties of
identification. To mitigate this problem, we take advantage of the fact that the
main sequence turn-off (MSTO) of the Sgr stream is much more well populated and
easily identifiable. The MSTO is not an ideal distance indicator itself, since
it exhibits considerable variation in absolute magnitude with differences in the
age and metallicity of the stellar population, but it does provide a reasonable
estimate of the stream's distance such that we can use it as an additional
constraint on the position of the RC. 

Our distance measurements are thus made with a two-step process, whereby we
first fit the bright edge of the MSTO, which unambiguously detects the Sgr
stream but does not yield a precise distance, then use that measurement to set
the range of apparent magnitudes which we use for detecting the red clump.  The
red clump fit then yields a precise distance.

Towards this goal, we first present an overview of the PS1 survey in
Section~\ref{sect_obs}, and present the PS1 view of the southern Sgr stream,
along with distance measurements, in Section~\ref{sect_view}.  We discuss the
results of those fits and their relation to other works in
Section~\ref{sect_results}, and conclude in Section~\ref{sect_conclusions}. 

\section{Observations and Data Processing}
\label{sect_obs}

\begin{figure*}
\epsscale{1.1}
\plotone{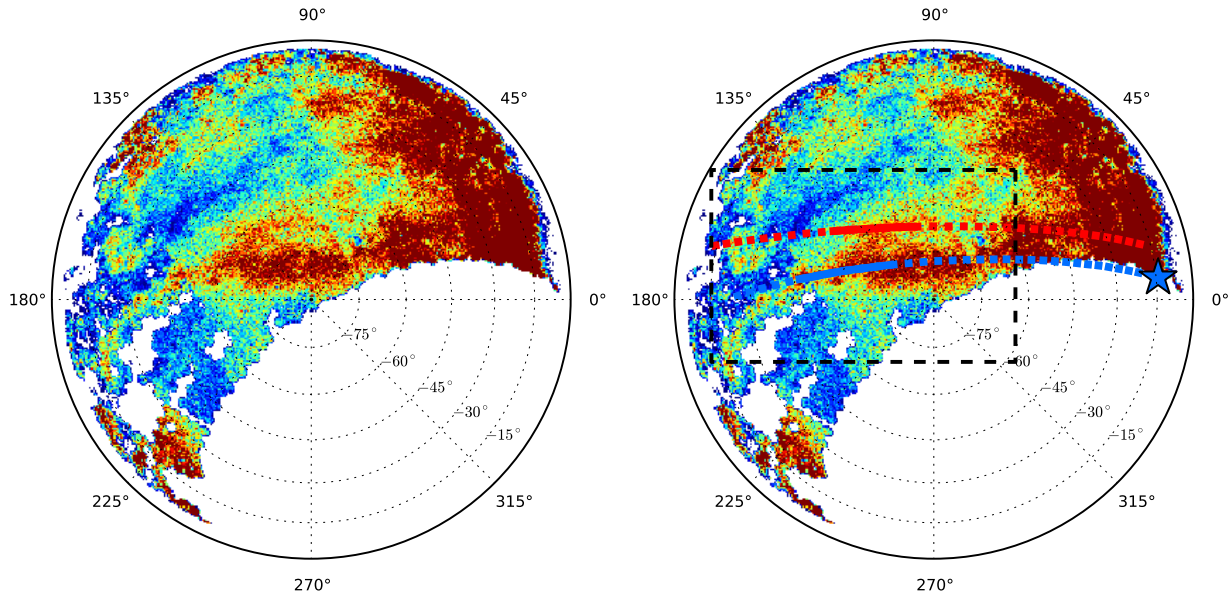}
\caption{Map of color-selected Sgr MSTO stars, shown in galactic coordinates and
centered on the south galactic cap. The right image shows the position of the
bright arm (solid blue) and the faint arm (solid red) as measured by
\citet{koposov12}, while the dashed versions show these lines extrapolated
beyond the SDSS coverage. The position of the dwarf is indicated by the star on
the right. The dashed rectangle indicates the region shown in
Figure~\ref{zoomed_image}.
\label{big_image}} 
\end{figure*}

The observations were conducted by the Panoramic Survey Telescope and Rapid
Response System 1 (Pan-STARRS1, hereafter referred to as PS1). PS1 is a 1.8m
telescope on Haleakela, Maui, which has been conducting a multi-faceted survey
since May 2010. The survey is designed both to detect transient, variable and
moving objects, such as supernovae, Kuiper belt objects, asteroids, and stellar
transits, and also to provide data to a number of static-sky projects, such as
large scale structure, galaxy properties, and Milky Way structure. The survey
covers the entire sky north of declination $-30^\circ$ (the so-called $3\pi$
survey), with 12 specific fields targeted for more frequent and deeper
observations (the medium-deep survey).

The pattern of observations on the sky is organized by a fixed tessellation
pattern which sets the boresight pointing locations. The timing of visits to
each position is set by the need to obtain an adequate baseline in time for
detection of transients and moving objects. The resulting set of observations
for each position is designed to comprise of four observations of ${\sim} 40$
seconds in each of the five PS1 filters, per observing season. The images are
processed by the Pan-STARRS1 Image Processing Pipeline
\citep[IPP,][]{magnier06}, which handles all image processing steps including
production of an object catalog for each image. Because of the computationally
intensive nature of the processing on even single images, the IPP does not
immediately produce stacked images. This is a necessary trade-off to enable the
IPP to process images rapidly after the observations were taken, which is
critical for the surveys of moving objects which require rapid follow-up
observations. 

As a result of this processing strategy, the currently available $3\pi$-survey
catalogs contain only the results of photometry performed on single images.
These catalogs are then cross-matched and merged to form an all-sky catalog. 
This does not increase the photometric depth in the same way that stacking
images would, so it is not possible to detect objects fainter than the limiting
magnitude of the deepest image. In order to reject spurious detections from
instrumental artifacts we require that objects be detected multiple times,
which makes our detection probability dependent on the number of visits to a
position on the sky.

The depth of the resulting catalog is therefore dependent on the number and
depth of individual exposures in each band in a particular region of the sky.
The number of visits to each position on the sky varies due to weather and
telescope downtime affecting the scheduling of visits. There are also variations
in observing conditions between images, since the survey is designed to take
data in marginal photometric conditions, which can affect the photometric
completeness at faint magnitudes. To assess this variation, we have
cross-matched the PS1 catalog with the SDSS ``stripe 82'' coadd catalog
\citep{annis11}. The coadded SDSS catalog is better than 90\% complete for stars
down to 23rd magnitude in g, r, and i bands, so we can assume that the any
non-detections of SDSS stars in the PS1 catalog is the result of PS1
incompleteness. From this cross-matching we can establish the 50\% completeness
magnitude over the range of observing conditions seen by the fields overlapping
stripe 82. In g-band the 50\% completeness ranges from approximately $\gps=21.4$
to $22.0$, in r-band from $\rps=21.2$ to $21.8$, and in i-band from $\ips=21.0$
to $21.8$. While these depths are fainter the range of magnitudes we use for in
this work, because the completeness is not a step function as a function of
magnitude there will be some variation in completeness even in the brighter data
we use for the MSTO fitting. From this cross-matching we can see that the
completeness at $\ips=21.0$ varies by approximately 20\% across the range of
observing conditions in stripe 82. This variation is reduced to 5\% or less at
magnitudes of $\ips=19.0$ or brighter. Much of this variation is likely to be
related to our ability to separate stars from galaxies, which is progressively
degraded under worse observing conditions. The photometric uncertainty on
detected objects is less variable and frequently very small, on average ranging
from 0.04 to 0.08 magnitudes near the MSTO (at approximately 21.5 mag in g and r
filters), and less than 0.03 mag at RC magnitudes (in r and i filters). 

One of the major challenges with any large survey is photometric calibration.
The initial calibration of the IPP catalog is performed by referencing stars in
each image to stars from the 2 Micron All Sky Survey \citep{skrutskie06}, which
provides a consistent way of calibrating single images across the entire sky.
However, now that the survey has obtained a sufficiently large set of
observations with overlapping images, it is possible to self-consistently
calibrate the survey using stars observed in multiple overlapping observations.
Commonly called ``\"{u}bercalibration'', this calibration strategy was adopted
by SDSS \citep{padmanabhan08} and has been implemented for PS1 by
\citet{schlafly12}. Comparisons between SDSS and the \"{u}bercalibrated PS1 data
show that the resulting measurements agree to $< 10$ mmag in the \gps, \rps, and
\ips filters \citep{schlafly12}. This calibration is applied to the data using
the Large Survey Database software (LSD, Juri\'c, in prep.), which provides a
fast and scalable interface to the large volume of data required. Analysis of
the calibrated data was also performed using LSD. All of the observations were
extinction corrected and de-reddened using the reddening maps of
\citet{schlegel98}. We make use of all PS1 $3\pi$ data obtained through January
18, 2012, at which time the survey has imaged the entire $3\pi$ target area,
though not always at the target photometric depth.

\section{The PS1 view of the Sgr stream}
\label{sect_view}


Figure~\ref{big_image} shows a map of stars in the southern Galactic hemisphere
with colors similar to that of the Sgr stream MSTO ($0.0 < (\gps - \rps)_0 <
0.4$ and $20.25 < g_{{\rm P1},0} < 22.10$). The Galactic center is on the right
edge, and the anticenter is on the left. The map is most sensitive to MSTO stars
at heliocentric distances between 25 and 45 kpc (assuming an old, metal poor
population). The position of the bright arm of the stream is illustrated by the
blue line on the right panel and the position of the faint arm \citep[as
measured by][]{koposov12} is shown by the red line. The stream is clearly
visible at $b < -60^\circ$, nearly passing through the south Galactic pole.
Towards the Galactic center, the stream skirts the edge of the PS1 coverage and
becomes lost in the higher density environment of the Galactic center. In the
anticenter direction, a combination of varying observation depth and the
increasing heliocentric distance of the stream makes it difficult to discern the
stream at $b>-45^\circ$. 

The Sgr stream in the southern hemisphere has also been seen in SDSS data
\citep{koposov12} over a somewhat smaller area. The SDSS and PS1 coverages are
shown in Figure~\ref{zoomed_image}. \citet{koposov12} have also shown that the
Sgr stream in the southern Galactic hemisphere is bifurcated in much the same
way as the northern part of the stream. They show that the density of MSTO stars
is asymmetric, with MSTO stars extending further perpendicular to the stream on
the northern side (top of Figure~\ref{zoomed_image}) than the southern side.
This is apparent in both the SDSS maps and the PS1 maps. 

\subsection{Distance Measurements}
\label{sect_dist}

\begin{figure}
\epsscale{1.20}
\plotone{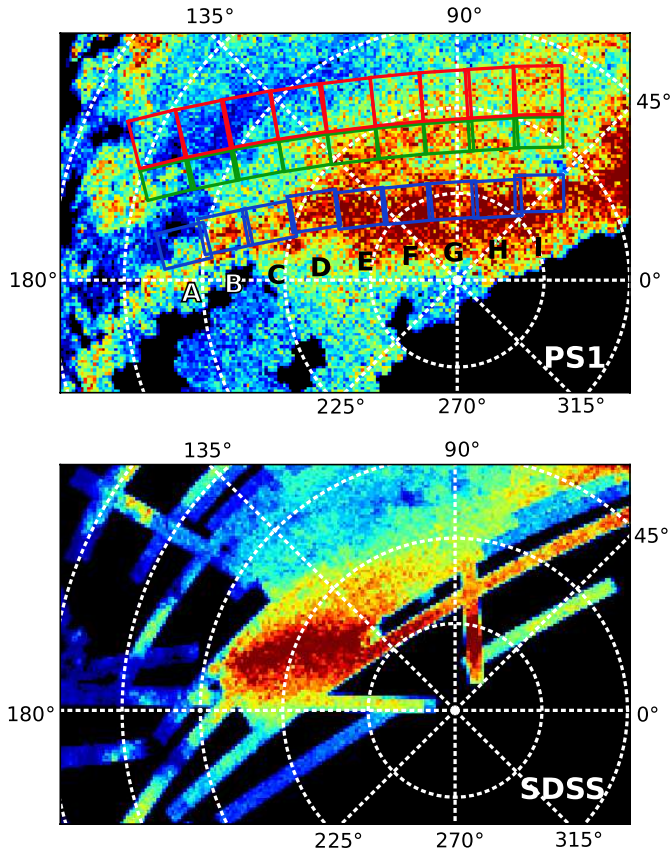}
\caption{Zoomed-in view of Figure~\ref{big_image}, focusing on the Sgr stream.
The top panel shows the data in PS1, while the bottom shows the same region in
the SDSS. The top panel also shows the regions targeting the bright arm of the
stream in blue, the faint arm regions in green, and the background regions in
red. The coordinate grid labels show Galactic longitude, and the concentric
rings are each separated by $15^\circ$ Galactic latitude. \label{zoomed_image}}
\end{figure}

We divide the length of the stream visible in the MSTO map into nine regions
spaced equally along the Sgr plane defined by \citet{majewski03}. Though this
plane was designed to fit the position of the stream on the sky, the peak of the
MSTO star density we observe lies slightly to the south of this plane. In order
to improve the signal to noise of our measurements, for each of the nine regions
we construct a histogram of MSTO star density as a function of distance off of
the Sgr plane, then center our target on-source regions on the peak of these
histograms. In general the region centers are approximately one degree south of
the Sgr plane, but each region is fit individually so there is some variation.
Not all of this variation is necessarily physical though, since we are also
affected by small gaps in the PS1 sky coverage. As a result, the positions of
these regions may only approximate the true position of the Sgr stream on the
sky. The on-source regions cover $6^\circ$ along the stream and
are $5^\circ$ wide perpendicular to the stream. The field centers are listed in
Table~\ref{rc_table}, and the on-source regions are shown in the top panel of 
Figure~\ref{zoomed_image} in blue.

In addition to placing on-source regions near the Sgr plane, we also placed
regions $17^\circ$ away from the plane to serve as background regions for the
MSTO fits. Ideally these background regions would be located at similar galactic
latitude to the target regions, but at high galactic latitude the PS1 sky
coverage does not include sufficient area that is not contaminated with the Sgr
stream itself for this to be possible. Instead we adopted the strategy of
background regions parallel to the stream, which for regions near the disk tends
to approximate a constant galactic latitude selection process due to the
position of the stream on the sky. At high galactic latitude this approximation
begins to break down, but the background contamination is much less significant
since those fields are far away from the disk. We also placed regions that
targeted the faint arm of the stream approximately $10^\circ$ away from the
plane, where the MSTO distribution shows a second peak or at least show signs of
being spatially extended to the north. The resulting regions are shown in the
top panel of Figure~\ref{zoomed_image}, with the faint arm regions in green and
the background regions in red.

\begin{figure}
\epsscale{1.2}
\plotone{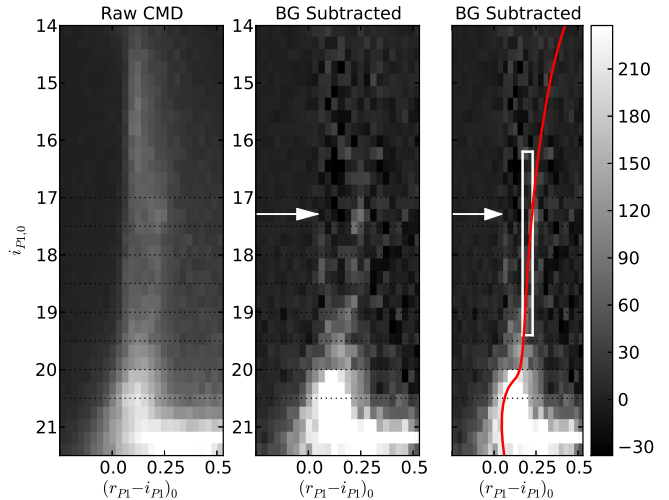}
\caption{CMDs for region E, showing the original
on-source CMD (left), the background-subtracted CMD (center), and the
background-subtracted CMD (right) with the RC color-cuts and an isochrone for an
11 Gyr old population with [Fe/H]$=-1.2$ overlaid \citep[from][]{marigo08}. The
horizontal arrow indicates the apparent magnitude of the red clump.
\label{hess}}
\end{figure}

A background-subtracted color-magnitude diagram (CMD) for a sample region is
shown in Figure~\ref{hess}. The background region was scaled to match the area
of the target region. An isochrone of an 11 Gyr old population with
[Fe/H]$=-1.2$ from \citet{marigo08} is shown for reference \citep[computed with
filter curves for the PS1 system;][]{tonry12}. This metallicity and age match
measurements of the stream in the north, though the stream exhibits a broad
range of metallicities \citep{chou07}. The Hess diagrams for the target fields
were used to determine the color cuts that would best select the RC ($0.17 <
(\rps-\ips)_0 < 0.23$) and the MSTO ($0.0 < (\rps-\ips)_0 < 0.10$). The RC color
selection is shown by tall white rectangle.

\begin{figure*}
\plotone{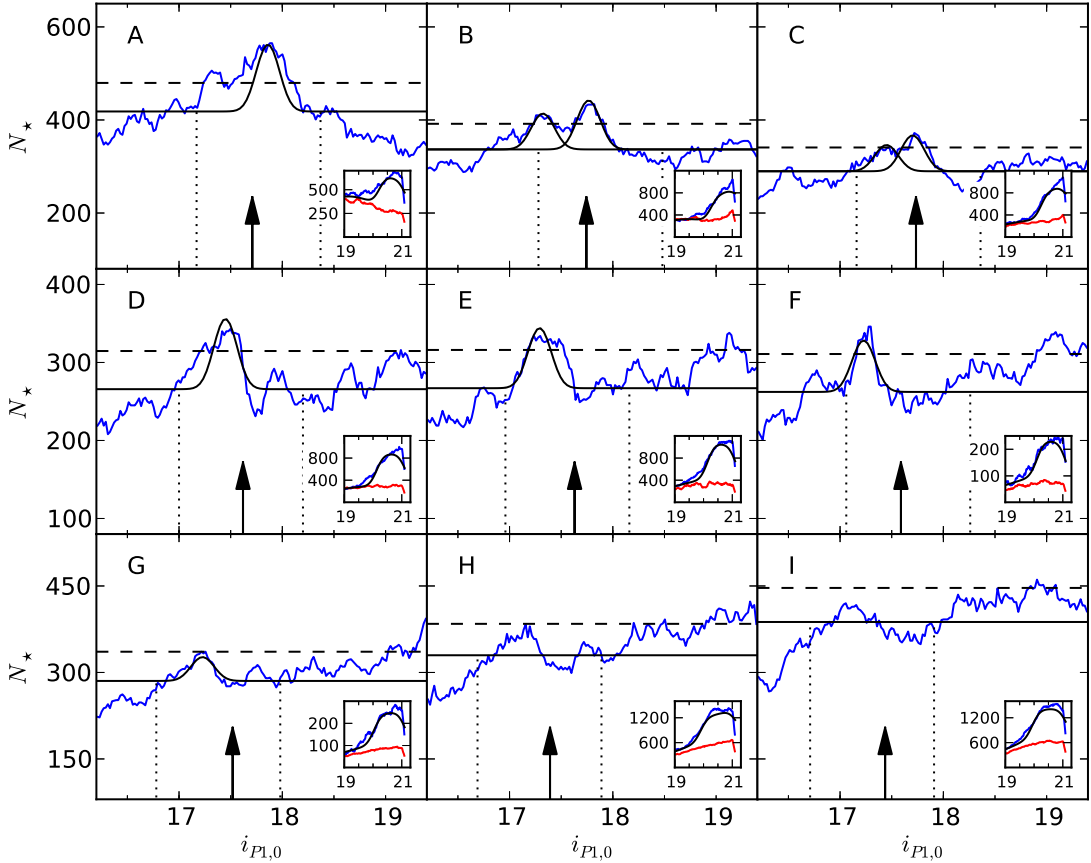}
\caption{Running histograms for the regions on the bright arm of the stream. The
blue line shows the running histogram, while the black lines show the best
fitting model (or multiple models in the case of multiple candidate peaks.) The
horizontal dashed line indicates the level at which an overdensity would have
a $3\sigma$ significance over the computed background level. The insets shows
the MSTO fits used as a prior on the RC fits, with the on-stream data shown in
blue, an off-stream background region shown in red, and the best fit model in
black. The arrow shows the position of the red clump suggested by the MSTO fit,
and the range of RC values allowed by the MSTO prior is shown by the vertical
dotted lines. In cases where there was no peak with significance greater than
2-$\sigma$, we do not plot a fit and instead denote the background level with a
solid horizontal line. \label{rc_fits}} 
\end{figure*}

\begin{figure*}
\plotone{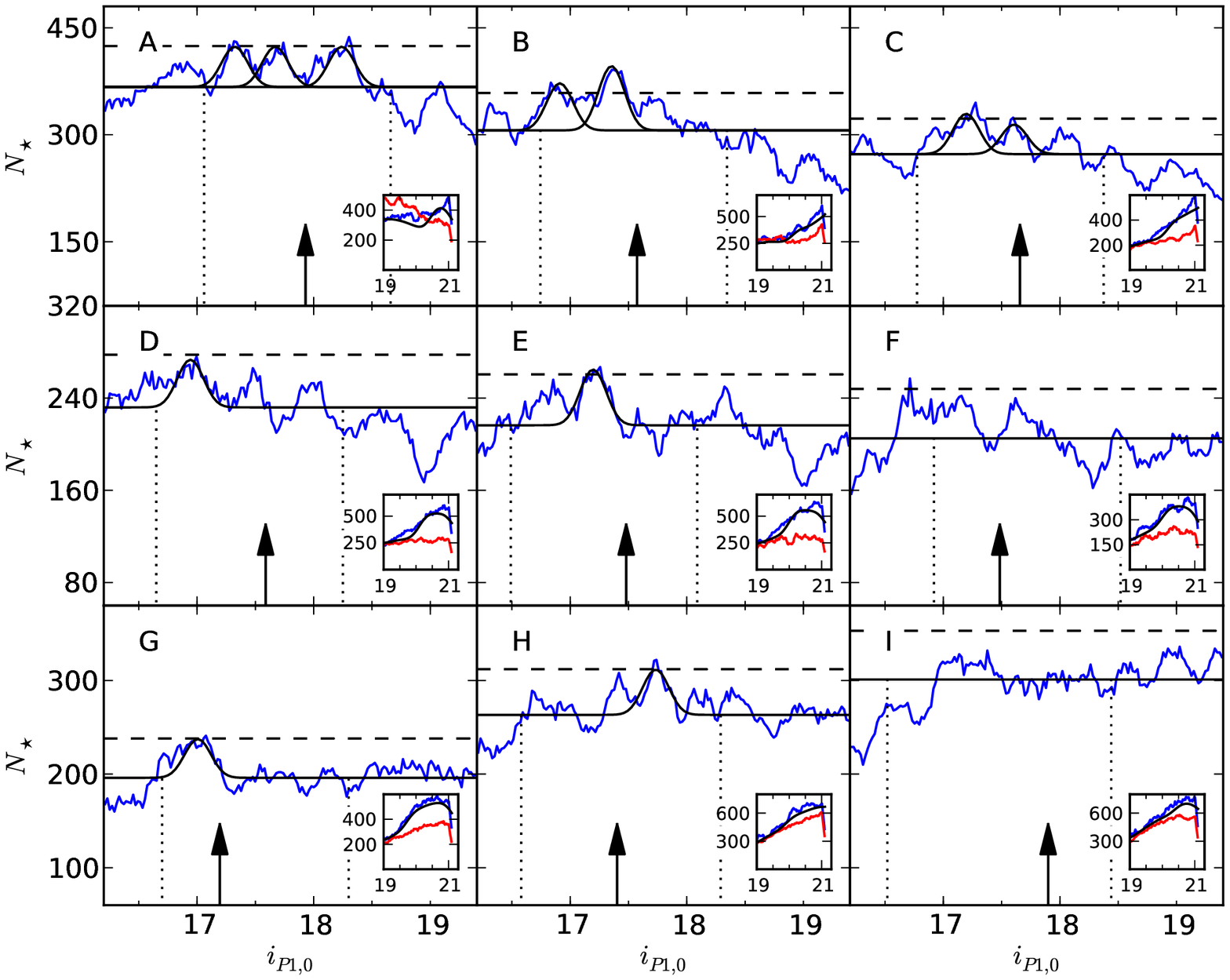}
\caption{RC fits for the regions placed on the faint arm of the stream. The
lines are the same as in Figure~\ref{rc_fits}. However, because we set the
window for the bifurcation fits with the measured RC position in the on-stream
regions, the allowed RC range (vertical dotted lines) may not align exactly with
the center position suggested by the MSTO fit (vertical arrow). 
\label{rc_bif}}
\end{figure*}

These color cuts were then used to construct histograms of the apparent
magnitudes of stars with the selected colors in each region. These regions are
shown in Figure~\ref{rc_fits} for the bright arm regions and Figure~\ref{rc_bif}
for the faint arm. For presentation purposes the figures show ``running
histograms'', similar to those used in \citet{correnti10} and
\citet{bellazzini05}. In these running histograms, each point is spaced by 0.02
magnitudes (the step size), but the value of each point is the number of stars
within a 0.2 magnitude bin centered on the point (the bin size). The procedure
is essentially a boxcar filter. The resulting step values are not statistically
independent of each other, but the running histogram has the advantage that the
visual appearance of the data is not affected by the particular bin positions in
the histogram. That is, if the RC happens to fall on the edge between two bins of
a conventional histogram, its signal appears more spread out than if it fell in
the center of a single bin. The bin width is also chosen so that it roughly
corresponds to the size of the RC, thus maximizing the signal to noise. The
running histograms are primarily a presentation tool; all of the statistical
fits which made use of histograms used ``normal'', fixed-bin histograms. 

\subsubsection{MSTO Fit}

After constructing the histograms of the MSTO and RC stars, we fit the
apparent magnitude of the MSTO using a maximum likelihood method, modeling the
edge of the MSTO as a step function superimposed on smooth background, as
measured from the off-stream field and fit by a third-order polynomial. The
step function was then convolved with a Gaussian with a width of $\sigma = 0.10$
magnitudes to account for both photometric scatter and intrinsic scatter in the
stellar luminosities. The result of these fits can be seen in the insets of
Figure~\ref{rc_fits}. In general these fits are very satisfactory, but it is
clear that the edge of the MSTO is not a very sharp transition; the transition
can span as much as half of a magnitude. 

To use the MSTO fit as a prior on the RC fit, we must determine the difference
in apparent magnitude between the two features. This calibration is inexact due
to the various complicating factors associated with the MSTO but still yields
useful results, since for this application we are only using it for different
parts of the same stream. Since the composition of stellar populations in the
stream is unlikely to vary greatly across the relatively modest length of stream
available to us, the RC-MSTO offset should also only exhibit small variations.
We computed this offset by creating synthetic CMDs from the \citet{marigo08}
isochrones (using a \citet{chabrier01} initial mass function), then applying the
same color cuts and using the same maximum likelihood fits for the MSTO and RC
position as were used on the PS1 data. Using an isochrone for an 11 Gyr old,
[Fe/H]$=-1.2$ population the RC-MSTO offset was computed to be 2.50 magnitudes
in i-band. Varying the metallicity of the isochrone from [Fe/H]$=-0.5$ to
[Fe/H]$=-1.3$ (roughly the range of stream metallicities seen in the northern
arm by \citet{chou07}) and also varying the age of isochrone from 6 Gyr to 11
Gyr, produced RC-MSTO offsets ranging from 2.71 through 2.40. The prior we adopt
on the RC fit is therefore broader than this range of variation, but centered on
an offset of 2.50 magnitudes.

\subsubsection{RC Fit}

A necessary step in fitting the RC is the determination of the smooth background
level, which by number dominates over the RC. As shown in the CMD
(Figure~\ref{hess}), the RC color cut we used is just slightly redder than the
region of the CMD populated by nearby halo stars, which significantly outnumber
the RC stars we are attempting to detect. Though the peak density of these stars
occurs slightly bluer in color than the RC, both intrinsic scatter in color and
photometric error causes some of these stars to fall into our color cuts.
The apparent magnitude distribution of stars in this region of the CMD is
approximately constant and not strongly affected by observing conditions. As a
result, we model the RC background as a third order polynomial in color (between
$(\rps-\ips)_0=0.15$ and $0.35$) and a constant in apparent magnitude. The polynomial
fitting excludes the RC color window to prevent biasing the background
estimation.

This approach to modeling the background has a number of advantages. The primary
reason for adopting this method is that the amount of background contamination
to the RC color cut is highly sensitive to photometric errors in the
observations. Greater observational uncertainties causes the large number of
MSTO-color stars to occupy a wider range of colors, causing more to fall into
the RC color cut and hence a higher background. This is particularly true for
PS1, which has data spanning a range of observing conditions. This makes it very
difficult to use off-stream regions as representative of the background level,
since it is practically impossible to find suitable background regions that
reproduce the observing conditions of an on-stream region. For this reason we
also do not attempt to fit a precise function to the background as a function of
apparent magnitude; though in several of the regions the background deviates
from flatness, it generally does not do so in a way that could be modeled by
comparison to an off-source region.  Additionally, since our RC fits are
constrained by the MSTO prior to a small range in apparent magnitude, it only
matters that our background estimation matches that small part of the histogram
well. The assumption of a constant background level is quite suitable under
these conditions.

After we have obtained the prior on the RC fit from the MSTO, we fit to the data
the sum of a Gaussian function with a fixed width of 0.1 magnitudes and the
constant background level. The MSTO fit and the RC-MSTO calibration were used to
set a flat prior within 0.6 mag of the center. The prior is broad enough to
account for any variation in the age and metallicity of the stream, along with
the inherent uncertainty and variability in the MSTO measurement as a result of
observational factors. The scale factor on the Gaussian is then marginalized
out, and the best fitting magnitude and uncertainty is found by taking moments
of the likelihood. In general this produces good fits, with the statistical
uncertainty in the measurement significantly smaller than the uncertainty of the
RC calibration. 

In a small number of fields, multiple distinct peaks appear in the histograms
with significance greater than $3\sigma$. In many of these cases one of the
peaks is more likely given the position of the stream in adjacent fields, but
for completeness we report both distances. In some other fields, the RC appears
to span a larger range of absolute magnitudes than expected, or exhibits some
asymmetry that may indicate some feature is present other than the ``pure'' Sgr
RC at a single distance. In these cases we do not have enough information to
report multiple distances, since it would be an ambiguous exercise to decompose
the noisy histogram into several RC components. Instead we report a single
distance but with an uncertainty that reflects the entire range of magnitudes
where the value of the running histogram exceeds the background by
$3\sigma$, as estimated from pure Poisson noise (corresponding to the horizontal
dashed lines in Figures~\ref{rc_fits} and \ref{rc_bif}). This effectively
determines the range of magnitudes at which, if our assumption of a constant
background holds, a statistically significant overdensity exists, even if we are
unable to conclusively identify its origin.

An additional source of uncertainty comes from our assumption of the faint arm's
position on the sky. In selecting our regions for the distance measurements we
have assumed that the faint arm parallels the bright arm over the region seen by
PS1, but our MSTO maps are unable to localize the faint arm precisely enough to
verify that this is the case. As shown by \citet{koposov12}, the streams may be
converging towards the Sgr dwarf at a rate of 0.05 degrees per degree along the
stream. Our target regions are designed to be sufficiently wide perpendicular to
the stream to encompass such gradients, but we have also tested our faint arm
fits with the target regions placed along paths converging or diverging from the
bright arm, with angles ranging from 0.10 deg/deg to -0.10 deg/deg. The results
still generally agree well with the results from the parallel-stream assumption,
but there is some variation in distances of approximately the same order or less
than the uncertainties on individual distance measurements. As a result we have
incorporated these additional uncertainties in our reported errors.

\begin{figure}
\epsscale{1.2}
\plotone{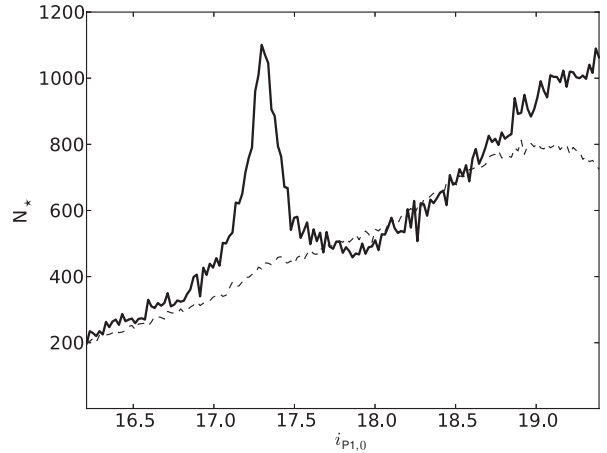}
\caption{Running histogram of RC color-selected stars in the main body of the
Sgr dwarf. The solid line shows the histogram from the dwarf, while the dashed
line shows the histogram for a background region. \label{dwarf_RC}}
\end{figure}

In order to use the RC as a distance indicator it is necessary to have a
calibration of the feature's absolute magnitude. This is complicated in the case
of the RC by the fact that its absolute magnitude is weakly dependent on the age
and metallicity of the stellar population \citep{seidel87,jimenez98}. These
variations can be minimized by using a red photometric passband
\citep{paczynski98} and by calibrating the RC magnitude with stars of the same
stellar population at a known distance. This ``differential'' measurement was
employed by \citet{correnti10} and is the method we use here. Since the Sgr
dwarf lies within the PS1 survey area we are able to directly use the survey's
observations of the main body of the dwarf to calibrate our RC distance
measurements along the stream. This avoids the need for any conversion between
photometric systems. The histogram of RC color-selected stars in the main body
of the dwarf is shown in Figure~\ref{dwarf_RC}, in which the RC is immediately
obvious. We fit the Sgr dwarf RC using the same methods as for the stream, but
with a polynomial background measured from a neighboring off-source region
instead of the constant used for the stream. The resulting fit measures the
apparent magnitude of the RC to be  $\ips=17.27 \pm 0.05$. We adopt the Sgr dwarf
distance measurement by \citet{monaco04}, which used the tip of the red giant
branch to measure a distance of $26.30 \pm 1.8$ kpc, or a distance modulus of
$(m-M)_0 = 17.10 \pm 0.15$ (but see \citet{kunder09} for a recent alternative
distance). All of our distance measurements are based on this
number, and so all of our reported distances will scale directly with any
revised value of the distance to the dwarf.

\section{Results \& Discussion}
\label{sect_results}

\begin{figure}
\epsscale{1.2}
\plotone{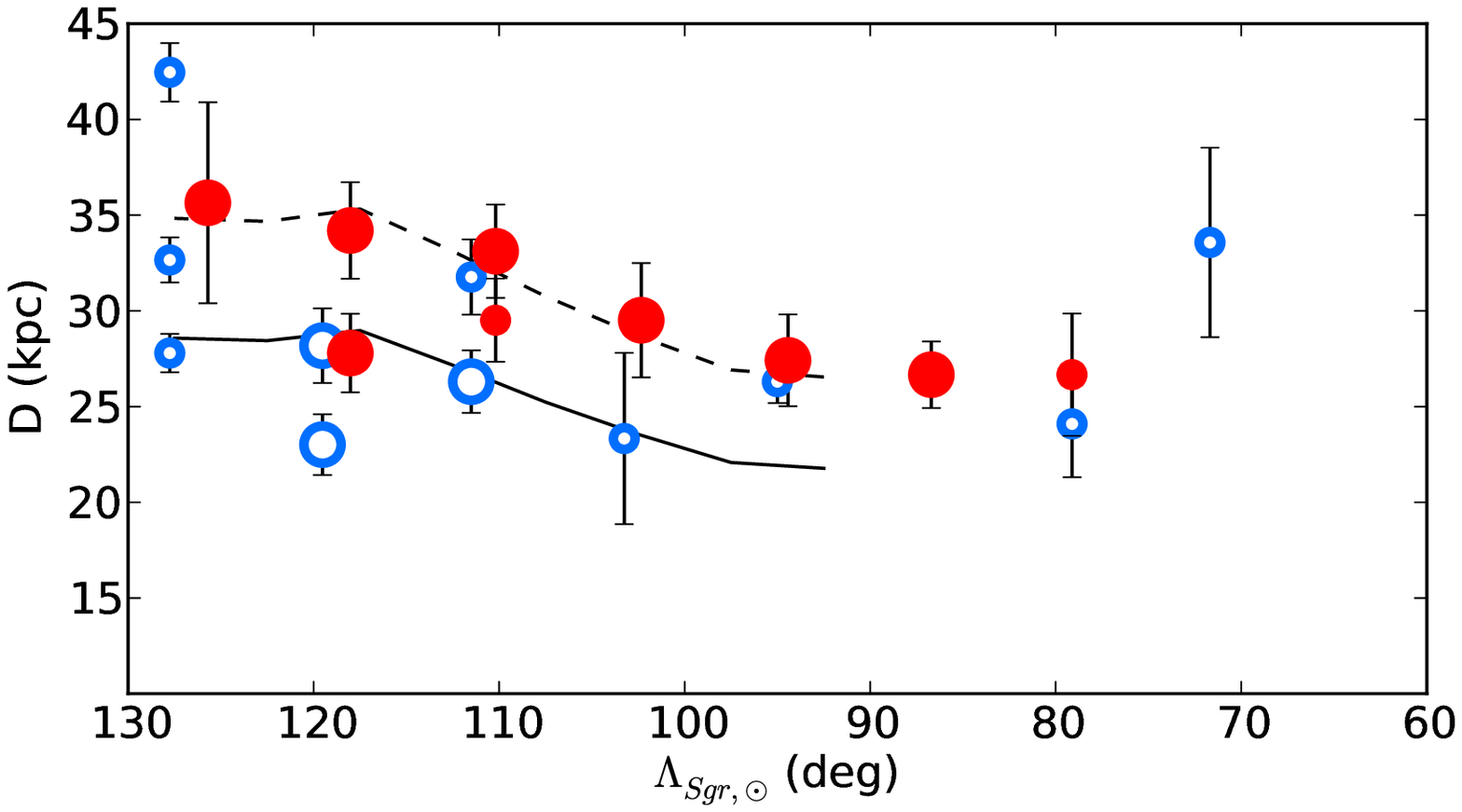}
\caption{Heliocentric distances to the bright arm (red, filled circles) and
faint arm regions (blue, open circles), plotted in the Sgr coordinate system of
\citet{majewski03}. The smaller points indicate detections with significances
between $2$ and $3\sigma$. The solid line indicates the distances to the Sgr
stream reported by \citet{koposov12}, while the dashed line shows the
measurements of \citet{koposov12} but adjusted to our calibration of the RC
absolute magnitude ($M_i = 0.17$). The Sgr dwarf is located at
$\Lambda_{Sgr,\odot} = 0$.
\label{distance_sgr}} 
\end{figure}

Our measured distances to the stream are shown in Figure~\ref{distance_sgr},
with the bright arm shown in red points and the faint arm in blue. While the
faint arm is just on the edge of detectability in several individual fields,
the consistent behavior of the faint arm across the length of the stream does
suggest that these detections do form a coherent structure. Despite the
significant uncertainties on each individual measurement of the distance, the
faint arm appears 3-5 kpc closer to the sun than the bright arm in several
different fields. A distance offset between the two arms
of the stream is also seen in the northern Galactic hemisphere, as seen by
\citet{yanny09}, \citet{niederste-ostholt10}, and \citet{ruhland11}, where the
faint arm (also called the ``B'' stream) also lies slightly closer to the sun
than the bright stream (the ``A'' stream). 

The distance to the main body of the trailing arm of the Sgr stream has also
been measured with the red clump in SDSS data by \citet{koposov12}. Their
distances are shown in Figure~\ref{distance_sgr} as the solid line. Comparing
their results to ours for the bright arm of the stream,
there is a clear difference of 7-10 kpc, which is consistent along the length of
the stream. This appears to match our measured distances to the faint arm, but
we believe that this is a coincidence. \citet{koposov12} show the density of RC
stars along the stream in SDSS as a function of apparent magnitude (their
Figure~5), and examination of this reveals that their detections of the RC lie
at nearly the exact same apparent magnitudes as ours. However, \citet{koposov12}
adopt an absolute magnitude of the RC as $M_i = 0.6$, citing
\citet{bellazzini06}, while we use $M_{\rm i,PS} = 0.17$ (the SDSS and PS1 i-band
filters are nearly identical). This discrepancy is significantly larger than the
uncertainty on the distance to the Sgr dwarf, which is 0.15 magnitudes. In
Figure~\ref{distance_sgr}, the dashed line shows the distance measurements of
\citet{koposov12} shifted by the difference between these calibrations, and the
resulting distances agree well with our data. It thus appears that the
discrepancy between the two results is caused by differences in the two RC
absolute magnitude calibrations.

Though \citet{koposov12} refers to \citet{bellazzini06} as the source of their
i-band RC calibration, it is unclear how this i-band measurement was obtained,
since \citet{bellazzini06} uses observations performed entirely in the B and V
bands and include no mention of the SDSS-i filter. With no further details
provided by \citet{koposov12} we cannot diagnose the process that lead to this
$M_i = 0.6$ calibration. We are confident that our own calibration is robust: we
detect the RC in the Sgr dwarf itself, using identical filters and observed as
part of the same survey as our detections of the stream, thus minimizing the
systematic uncertainty associated with color transformations and comparisons
between data of different origins. One other possible source of variation
between calibrations is the assumed extinction values to the Sgr dwarf, since it
lies close to the Galactic plane, but it is unlikely that extinction could
account for the magnitude of the calibration difference. \citet{bellazzini06}
computes an average reddening of E(B-V)=0.116, while we use the value from the
maps of \citet{schlegel98} which report E(B-V)=0.15. In terms of extinction this
amounts to a difference in $A_i$ of 0.07 magnitudes, much less than the
difference in the two calibrations. 

\begin{figure*}
\epsscale{1.2}
\plotone{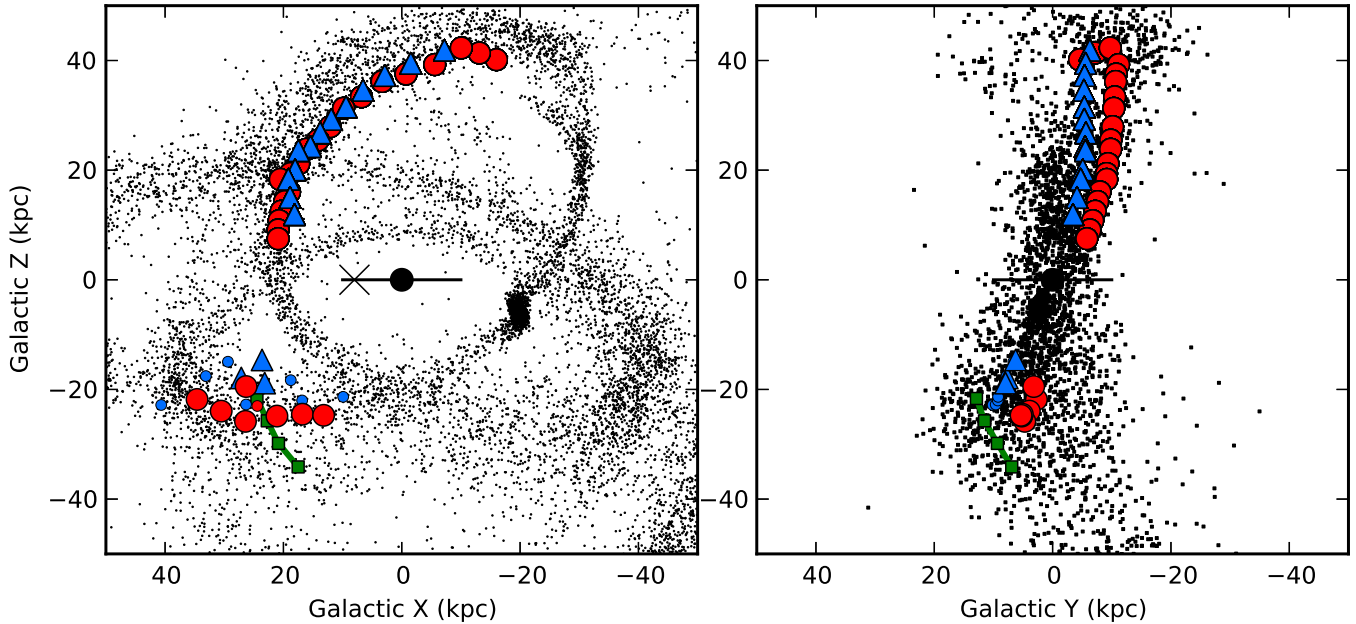}
\caption{Detected positions of the Sgr stream projected onto the Galactic X-Z
plane (left), which approximates the orbital plane of the stream, and in the Y-Z
plane (right), which is perpendicular to the orbit. The position of the Galactic
disk is shown by the horizontal line, and the sun is marked with an $\times$.
In the south, red circles and blue triangles signs indicate the position of
statistically significant detections of the RC on the bright arm and the faint
arm fields, respectively. The smaller points indicate detections with
significances between $2$ and $3\sigma$. In the northern hemisphere, SDSS
detections of the stream from \citet{niederste-ostholt10} are shown with red
circles for the bright arm, and blue triangles again for the faint arm.  The
position of the Cetus stream \citep{newberg09} is also shown by the green
squares. The black dots are data from the simulations of \citet{law10}. 
\label{sgr_XYZ}}
\end{figure*}

In Figure~\ref{sgr_XYZ}, the measured distances are compared to the numerical
model of the Sgr tidal streams produced by \citet{law10}. The figure shows a
projection of the stream onto the Galactic X-Z plane, which approximates the
plane of the Sgr stream, and the Y-Z plane. The agreement between our distances
and the position of the trailing (southern) arm of the Sgr debris is excellent.
This is in light of the fact that the simulation was not selected to match the
distance to the stream, but only the measured radial velocities to the leading
and trailing streams. In the right panel of Figure~\ref{sgr_XYZ}, the Sgr
dwarf appears slightly offset from the plane of the bright arm of the stream,
but this is the result of a projection effect. Because the Sgr stream does not
lie exactly in the X-Z plane, the projection of the stream onto the Y-Z plane
will exhibit some variation in the Y direction. Because the dwarf lies on the
opposite side of the Galactic center from the majority of the stream points,
they will appear to be offset in Y when projected in this manner.

Figure~\ref{sgr_XYZ} also shows the position of the Cetus stream, as measured by
\citet{newberg09}. Though the stream crosses the position of the Sgr stream on
the sky, and in the X-Z plane of Figure~\ref{sgr_XYZ} (left panel) appears to
be coincident with detections of the faint arm, the X-Z projection of
Figure~\ref{sgr_XYZ} clearly shows that the Cetus stream passes behind the Sgr
stream and not through it. It is therefore unlikely that our detections are
confused with the Cetus structure.

Though there is currently no model for the origin of the northern and southern
bifurcations that matches all of the available data, it is still instructive to
examine how our results compare to the existing models in order to develop a
sense of how a plausible bifurcation may behave in simulations. One hypothesis
is that the Sgr dwarf was originally a disc galaxy, and that the material in the
faint arm was stripped during a different pericentric passage than the bright
arm \citep{penarrubia10}. This theory is now thought to be untenable, since
the Sgr dwarf is observed to have very little residual rotation
\citep{penarrubia11}, but the simulations used in \citet{penarrubia10} are still
useful for assessing the difficulty of reproducing the southern bifurcation
simultaneously with the north. In the scenario that best reproduced the northern
bifurcation, where the two streams are nearly coincident, the material from the
two streams in the southern hemisphere is widely divergent. Material stripped in
the same pericentric passage as the northern faint arm rapidly increases in
helicentric distance towards the Galactic anticenter (from roughly 25kpc to
80kpc), while material from the bright arm stays at relatively constant
heliocentric distance (from approximately 20kpc to 10kpc). These simulations
were not designed to reproduce the stream's behavior in the southern hemisphere
and cannot be faulted for not doing so, but the discrepancy highlights the
difficulty inherent in the problem. 

\begin{table*}
\begin{tabular}{lccccccccc}
\tableline\tableline
Field &  A & B & C & D & E & F & G & H & I  \\
Field center (RA,Dec) &  (50,8) & (43,5) & (37,1) & (29,-2) & (23,-7) & (16,-11) & (9,-14) & (3,-18) & (356,-20)  \\
Field center ($\Lambda_{sgr,\odot}$,$B_{sgr,\odot}$) &  (126,-1) & (118,-1) & (110,-1) & (102,0) & (94,-0) & (87,-1) & (79,-1) & (72,-1) & (65,-0)  \\
Distance (kpc) &  35.6 & 34.2,27.8 & 33.1,29.5\** & 29.5 & 27.4 & 26.7 & 26.7\** &  &   \\
\tableline
Bifurcation center (RA,Dec) &  (47,18) & (40,14) & (32,11) & (25,7) & (18,3) & (11,-1) & (4,-5) & (358,-8) & (351,-11)  \\
Bifurcation center ($\Lambda_{sgr,\odot}$,$B_{sgr,\odot}$) &  (128,9) & (120,9) & (112,10) & (103,10) & (95,10) & (87,10) & (79,10) & (72,10) & (64,10)  \\
Bifurcation Distance (kpc) &  42.5\**,27.8\**,32.7\** & 28.2,23.0 & 26.3,31.8\** & 23.3\** & 26.3\** &  & 24.1\** & 33.6\** &   \\
\tableline
\end{tabular}
\caption{
Positions of the regions along the Sgr stream used to detect the RC, along with the
corresponding distance measurements. In cases where multiple overdensities could
correspond to the red clump, the distances from each candidate position are
listed. Detections with signifiance between $2$ and $3-\sigma$ are marked with
an asterisk.
\label{rc_table}
}
\end{table*}

\section{Conclusions}
\label{sect_conclusions}

In this work we have used data from Pan-STARRS1 to show the spatial extent of
the Sgr stream over $60^\circ$ of its orbit in the southern Galactic hemisphere.
The position we observe of the stream on the sky matches well with observations
from other surveys \citep{koposov12} and with simulations \citep{law10}. We have
also shown that the stream in the southern hemisphere exhibits a bifurcation
similar to that seen in the northern hemisphere, as was reported by
\citet{koposov12}. 

Using a combination of the MSTO and the RC as distance indicators, we have
measured the distance to both arms of the southern stream. Our results for the
stream again agree well with the simulations of \citet{law10}, but disagree with
the distances measured by \citet{koposov12}. We believe that this disagreement
is the result of differences in calibrations of the RC absolute magnitude, and
not differences between the SDSS and PS1 surveys, but it is not possible for us
to further deduce the exact cause of the disagreement. Our calibration strategy,
based on direct observations of the Sgr dwarf within the PS1 survey, gives us
confidence that we have one of the most direct calibrations possible, free of
any photometric transforms between data sources. Our distance measurements
consistently place faint arm of the stream slightly closer in heliocentric
distance than the bright arm. This is similar to the behavior seen in the
northern hemisphere, again suggesting that the bifurcation is the result of some
intrinsic behavior in the accreting Sgr system and not a coincidental overlap of
multiple wraps.

Unfortunately there is no model for the bifurcation that can adequately
explain the observed behavior. Models requiring multiple wraps of the stream
\citep{fellhauer06} or internal rotation of the Sgr dwarf appeared to be
untenable even prior to the discovery of the bifurcation in the south
\citep{yanny09,niederste-ostholt10}. It has been speculatively suggested by
\citet{koposov12} that what we call the Sgr stream could be the result of the
accretion of a pair of dwarf galaxies simultaneously. Our observations should
provide a quantitative basis against which simulations of a double accretion
scenario can be tested. It is remarkable that one of the most prominent and most
well-studied tidal streams has proven to be the most elusive streams to explain.
While our observations can at the moment only deepen the mystery, we are hopeful
that the added information on the stream's behavior will be used to properly
explain the bifurcation in future efforts.

\acknowledgments

We would like to thank the anonymous referee for their careful reading and
helpful comments. This work was partially supported by NSF grant AST 1008342.
NFM and EPM were both partially funded by Sonderforschungsbereich SFB 881 ``The
Milky Way System'' (subproject A3) of the German Research Foundation (DFG).

The Pan-STARRS1 Survey has been made possible through contributions of the
Institute for Astronomy, the University of Hawaii, the Pan-STARRS Project
Office, the Max-Planck Society and its participating institutes, the Max Planck
Institute for Astronomy, Heidelberg and the Max Planck Institute for
Extraterrestrial Physics, Garching, The Johns Hopkins University, Durham
University, the University of Edinburgh, Queen’s University Belfast, the
Harvard-Smithsonian Center for Astrophysics, and the Las Cumbres Observatory
Global Telescope Network, Incorporated, the National Central University of
Taiwan, and the National Aeronautics and Space Administration under Grant No.
NNX08AR22G issued through the Planetary Science Division of the NASA Science
Mission Directorate.

{\it Facilities:} \facility{Pan-STARRS1}.

\clearpage
\end{document}